\documentclass[12pt,draftcls,onecolumn,english]{IEEEtran}
\usepackage{color}

\usepackage{xtab}
\usepackage{scrextend}
\usepackage{cite}

\usepackage{amsthm}

\usepackage{amsmath,amsfonts, amssymb}

\usepackage{bbm}
\usepackage{bm}
\allowdisplaybreaks
\usepackage{subcaption}
\usepackage{graphicx}
\graphicspath{{./fig/}}

\usepackage{algorithm,algorithmic}%

\usepackage{url}

\usepackage{multirow}

\newcommand{\RV}[1]{{\textcolor{black}{{#1}}} }

\theoremstyle{plain}

\theoremstyle{definition}

\theoremstyle{remark}

\title{Probabilistic Hosting Capacity Analysis via Bayesian Optimization}

\author{Xinbo Geng, Lang Tong, Anirban Bhattacharya, Bani Mallick, and Le Xie
\thanks{
X. Geng and L. Tong are with the School of Electrical and Computer Engineering, Cornell University, Ithaca, NY, 14853.
Emails: \{xg72, lt35\}@cornell.edu.
X. Geng, A. Bhattacharya, B. Mallick, and L. Xie are affiliated with Texas A\&M Research Institute for Foundations of Interdisciplinary Data Science (FIDS), College Station, TX, 77853. Emails: \{xbgeng, le.xie\}@tamu.edu and \{anirbanb, bmallick\}@stat.tamu.edu. Correspondence is suggested to be sent to Le Xie. 
This work is supported in part by NSF CCF-1816397 and ECCS-1809830, and in part by NSF CCF-1934904, ECCS-2038963 and ECCS-1839616.
}
}

\begin{document}
\maketitle

\begin{abstract}
This paper studies the probabilistic hosting capacity
analysis (PHCA) problem in distribution networks considering
uncertainties from distributed energy resources (DERs) and
residential loads. PHCA aims to compute the
hosting capacity, which is defined as the maximal level of DERs
that can be securely integrated into a distribution network while satisfying operational constraints with high probability. 
We formulate PHCA as a chance-constrained optimization problem, and model the uncertainties from DERs and loads using historical data. \RV{Due to non-convexities and a substantial number of historical scenarios being used, PHCA is often formulated as large-scale nonlinear optimization problem, thus computationally intractable to solve.} To address the core computational challenges, we propose a fast and extensible framework to solve PHCA based on Bayesian Optimization (BayesOpt). Comparing with state-of-the-art algorithms such as interior point and active set, numerical results show that the proposed BayesOpt approach is able to find better solutions (25\% higher hosting capacity) with 70\% savings in computation time \RV{on average.}
\end{abstract}

%
\IEEEpeerreviewmaketitle

\section{Introduction} 
\label{sec:introduction}
The rapidly growing distributed energy resources (DERs) are reshaping the design and operation of distribution power networks.
The excessive amount of reverse power flow and intermittent DERs gives rise to an array of operational risks such as overloading of transformers and feeders, protection failures, over and under voltages, excessive line losses, and high harmonic distortion \cite{ismael_state---art_2019}.
\RV{To avoid compromising the operational security and reliability,} distribution system operators (DSOs) often perform hosting capacity analysis (HCA) to compute \emph{hosting capacity},
which is defined as the amount of  production that can be integrated into a distribution network, above which the system performance becomes unacceptable \cite{ismael_state---art_2019,bollen_integration_2011}. HCA reveals the physical limits and major limiting factors of a distribution network to accommodate deep penetration of DERs. By controlling existing devices, it is shown that a distribution network can host more DERs thus defer costly planning decisions such as installing additional power lines and upgrading transformers.




One commonly accepted estimation on the hosting capacity of a distribution feeder is a fixed percentage (e.g., 15\%) of annual peak load as most recently measured at the substation \cite{ferc_small_2020}. However, this simplistic estimation is problematic as peak loads and hosting capacities are shown to be poorly correlated\cite{baldenko_determination_2016}. An alternative approach is to perform HCA and compute the hosting capacity for a given DER generation and load profile, e.g., \cite{capitanescu_assessing_2014}. This approach considers more factors such as network topology and spatial and temporal couplings, but it fails to capture the significant level of stochasticity brought about by DERs, which is the main limiting factor of deep DER integration. 

\RV{To accommodate the stochastic nature of DERs, various techniques of hosting capacity analysis have been proposed. These techniques can be categorized into three groups based on modeling of DER uncertainties: (1) using probability distributions \cite{al-saadi_probabilistic_2017,abad_probabilistic_2018}, (2) stochastic programming based approaches \cite{santos_new_2016-1,xu_enhancing_2019}, and (3) set-based description of DER uncertainties using robust optimization \cite{chen_data-driven_2017,chen_robust_2017-1}.}
\RV{Although different models of DER uncertainties lead to distinct HCA formulations, there are two things in common. First, most of them model uncertainties using a large number of sampled or historical data, e.g., \cite{santos_new_2016-1,xu_enhancing_2019,chen_data-driven_2017,chen_robust_2017-1}. Second, most of them quantify the operational risk of integrating DERs using the probability of violating operational constraints, e.g., \cite{al-saadi_probabilistic_2017,abad_probabilistic_2018,xu_enhancing_2019,chen_data-driven_2017}.
Motivated by this two common features, this paper studies the data-driven probabilistic hosting capacity analysis problem, which models the DER and load uncertainties using historical data, and restricting the probability of constraint violation within acceptable level. 
}

Due to the large number of scenarios and nonlinear power model equations, data-driven PHCA often requires solving large-scale non-convex optimization problems.
This paper addresses the computational challenges in PHCA and proposes an efficient and extensible computational framework.
Instead of solving non-convex optimization problems involving a large number of variables and scenarios, we formulate PHCA as a small-scale nonlinear optimization problem which only includes nodal hosting capacities as decision variables. The proposed PHCA formulation features a non-convex and computationally expensive constraint. This constraint is \emph{expensive} in the sense of evaluating it involves checking the feasibility of (optimal) power flow problems in every DER and load scenario. We use Bayesian Optimization to search for the \emph{global} optimal solution to PHCA while using as few evaluations of the expensive constraint as possible. 

\RV{The main contribution of this paper is an efficient and flexible computational framework for PHCA. Numerical results on the 56-node South California Edison distribution network show that the proposed BayesOpt approach is able to find better solutions (25\% higher hosting capacity) with 70\% savings in computation time. It is also worth mentioning that this paper might be the first attempt to solve power system optimization problems directly via BayesOpt, instead of using BayesOpt indirectly such as tuning hyperparameters of machine learning models.}

The notations in this paper are standard. All matrices and vectors are in the real field $\mathbb{R}$.
Matrices and vectors are in bold fonts, e.g., $\bm{A}$ and $\bm{b}$. The all-1 (all-0) vector of appropriate size is denoted by $\bm{1}$ ($\bm{0}$). The indicator function is $\mathbbm{1}(\cdot)$. The transpose of a vector $\bm{a}$ is $\bm{a}^\intercal$, the diagonal matrix formed using vector $\bm{a}$ is $\text{diag}(\bm{a})$, and the $j$th entry of vector $\bm{a}$ is $a_j$.
Sets are in calligraphy fonts. The cardinality of a set $\mathcal{S}$ is $|\mathcal{S}|$.
The upper and lower bounds on variable $\bm{v}$ are denoted by $\overline{\bm{v}}$ and $\underline{\bm{v}}$, respectively. 

The remainder of this paper is organized as follows. Section \ref{sec:hosting_capacity_maximization} introduces PHCA and conventional approaches to solve it. Section \ref{sec:hosting_capacity_maximization_via_bayesian_optimization} proposes a computational framework to solve PHCA based on Bayesian Optimization. Numerical results and discussions are in Section \ref{sec:case_study}.



\section{Hosting Capacity Analysis} 
\label{sec:hosting_capacity_maximization}

\subsection{DistFlow Model} 
\label{sub:distflow_model}
We study a distribution network $\mathcal{N} = (\tilde{\mathcal{V}}, \mathcal{E})$ with vertices $\tilde{\mathcal{V}} = \mathcal{V} \cup \{0\}$ and edges $\mathcal{E}\subseteq \tilde{\mathcal{V}} \times \tilde{\mathcal{V}}$. The substation node is denoted by $0$ and all other nodes are represented by $\mathcal{V}$. Nodal real and reactive \emph{injections} are denoted by $\bm{p} \in \mathbb{R}^{|\mathcal{V}|}$ and $\bm{q} \in \mathbb{R}^{|\mathcal{V}|}$. Nodal squared voltage magnitudes are $\bm{v} \in \mathbb{R}^{|\mathcal{V}|}$. The real and reactive flows on distribution lines are $\bm{P} \in \mathbb{R}^{|\mathcal{E}|}$ and $\bm{Q} \in \mathbb{R}^{|\mathcal{E}|}$. $P_{ij}$ and $Q_{ij}$ denote the real and reactive power flow on line $(i,j)$.
If the distribution network $\mathcal{N}$ is a tree (thus $|\mathcal{E}| = |\mathcal{\tilde{V}}|-1 = |\mathcal{V}|$), then the following \emph{DistFlow} model holds true \cite{low_convex_2014-1}. 
\begin{subequations}
\label{eqn:DistFlow}
\begin{align}
& p_j + P_{ij} = r_{ij} l_{ij} + \sum_{k: (j,k) \in \mathcal{E}} P_{jk} ,\\
& q_j + Q_{ij} = x_{ij} l_{ij} + \sum_{k: (j,k) \in \mathcal{E}} Q_{jk} ,\\
& v_i - v_j = 2(r_{ij} P_{ij} + x_{ij} Q_{ij}) + (r_{ij}^2 + x_{ij}^2) l_{ij}, \\
& l_{ij} = \frac{P_{ij}^2 + Q_{ij}^2}{v_i}, \\
& i \in \mathcal{V}\cup \{0\},~j \in \mathcal{V},~(i,j) \in \mathcal{E}. \nonumber 
\end{align}
\end{subequations}
For simplicity, the DistFlow model \eqref{eqn:DistFlow} in the remainder of this paper is denoted by \eqref{eqn:correspondence_DistFlow}.
\begin{equation}
\label{eqn:correspondence_DistFlow}
\big[\bm{P}, \bm{Q}, \bm{v} \big] = \text{DistFlow}(\bm{p}, \bm{q})
\end{equation}


\subsection{Probabilistic Hosting Capacity Evaluation} 
\label{sub:hosting_capacity_evaluation}
Given candidate locations $\mathcal{L} \subseteq \mathcal{V}$ of DERs, the (deterministic) hosting capacity evaluation problem is to check the feasibility of equation \eqref{eqn:hosting_capacity_passive_constraints} for multiple snapshots $t \in \mathcal{T}$.
\begin{subequations}
\label{eqn:hosting_capacity_passive_constraints}
\begin{align}
& \big[\bm{P}[t], \bm{Q}[t], \bm{v}[t] \big] = \text{DistFlow}(\bm{p}[t], \bm{q}[t]),&~t \in \mathcal{T}. \label{eqn:hosting_capacity_passive_constraints_DistFlow} \\
& \bm{p}[t] = \bm{A}^{\mathcal{L}} \text{diag}(\bm{\alpha}[t])\bm{\psi}  - \bm{d}[t],&~t \in \mathcal{T}. \label{eqn:hosting_capacity_passive_constraints_real} \\
& \bm{q}[t] = \bm{A}^{\mathcal{L}} \text{diag}(\bm{\eta})\text{diag}(\bm{\alpha}[t])\bm{\psi} - \bm{e}[t],&~t \in \mathcal{T}. \label{eqn:hosting_capacity_passive_constraints_reactive}\\
& \underline{\bm{v}} \le \bm{v}[t] \le \overline{\bm{v}},&~t \in \mathcal{T}. \label{eqn:hosting_capacity_passive_constraints_voltage} \\
& (P_{ij}[t])^2 + (Q_{ij}[t])^2 \le (\overline{S}_{ij})^2,~(i,j)~\in \mathcal{E},&~t \in \mathcal{T}. \label{eqn:hosting_capacity_passive_constraints_line}
\end{align}
\end{subequations}
The key elements in equation \eqref{eqn:hosting_capacity_passive_constraints} include: a given DER installation scenario $\bm{\psi} \in \mathbb{R}^{|\mathcal{L}|}$, a given DER generation profile $\{\bm{\alpha}[t]\}_{t \in \mathcal{T}}$, and real/reactive load profiles $\{\bm{d}[t], \bm{e}[t]\}_{t \in \mathcal{T}}$. Matrix $\mathbf{A}^{\mathcal{L}} \in \{0,1\}^{|\mathcal{V}| \times |\mathcal{L}|}$ is the DER location-bus adjacency\footnote{$\mathbf{A}_{ij}^{\mathcal{L}}=1$ if the $j$th DER is at bus $i$; $\mathbf{A}_{ij}^{\mathcal{L}}=0$ otherwise.} matrix. We assume that DERs operate in the maximum power point tracking (MPPT) mode and maintain fixed power factor $\bm{\eta} \in \mathbb{R}^{|\mathcal{L}|}$ by simple reactive power control.

Constraints \eqref{eqn:hosting_capacity_passive_constraints} consist of DistFlow equations \eqref{eqn:hosting_capacity_passive_constraints_DistFlow}, nodal real/reactive power balance \eqref{eqn:hosting_capacity_passive_constraints_real}-\eqref{eqn:hosting_capacity_passive_constraints_reactive}, voltage magnitude \eqref{eqn:hosting_capacity_passive_constraints_voltage} and line flow limits \eqref{eqn:hosting_capacity_passive_constraints_line}. \RV{To focus on the computational aspect,} this paper considers passive distribution networks. The problem formulation and proposed computational framework in Section \ref{sec:hosting_capacity_maximization_via_bayesian_optimization} can be easily extended towards more complicated constraints and active distribution network (ADN) settings. Detailed discussions are in Section \ref{sub:extensions}.

Hosting capacity evaluation checks if a DER installation scenario $\bm{\psi}$ is feasible for a particular DER and load profile. The main drawback of hosting capacity evaluation is the failure of capturing the stochastic nature of DERs. \emph{Probabilistic} hosting capacity evaluation takes the DER and load uncertainties into consideration and calculates the probability of violating operational constraints \cite{al-saadi_probabilistic_2017,abad_probabilistic_2018}.
Let constraints \eqref{eqn:hosting_capacity_passive_constraints} be represented succinctly as $\bm{g}(\bm{\alpha}, \bm{d}, \bm{e}; \bm{\psi} ) \le 0$. Given a DER installation scenario $\bm{\psi}$, probabilistic hosting capacity evaluation calculates the probability $\epsilon(\bm{\psi})$ of constraint violation.
\begin{equation}
\epsilon(\bm{\psi}) := 1-\mathbb{P}\left( \bm{g}(\bm{\alpha}, \bm{d}, \bm{e}; \bm{\psi} ) \le 0 \right)
\end{equation}


\subsection{Probabilistic Hosting Capacity Analysis (PHCA)} 
\label{sub:hosting_capacity_maximization}
This paper studies the probabilistic hosting capacity analysis problem, which is closely related with probabilistic hosting capacity evaluation in the previous section.
\begin{subequations}
\label{opt:CC-PHCA}
\begin{align}
\max_{\bm{0} \le \bm{\psi} \le \overline{\bm{\psi}}}~& \bm{1}^\intercal \bm{\psi}  \label{opt:CC-PHCA-objective} \\
\text{s.t.}~& \epsilon(\bm{\psi}) \le \overline{\epsilon} \label{opt:CC-PHCA-chance}
\end{align}
\end{subequations}
The objective of PHCA \eqref{opt:CC-PHCA-objective} is to identify (maximal) \emph{hosting capacity}, which is defined as the maximal level of DERs that can be securely integrated into a distribution network while restricting the risk of constraint violation $\epsilon(\bm{\psi})$ within acceptable level $\overline{\epsilon}$. Constraint \eqref{opt:CC-PHCA-chance} is a chance constraint, it ensures the operational constraint $\bm{g}(\bm{\alpha}, \bm{d}, \bm{e}; \bm{\psi} ) \le 0$ to be satisfied with probability at least $1-\overline{\epsilon}$ in the presence of DER and load uncertainties. The optimal solution $\bm{\psi}^*$ to \eqref{opt:CC-PHCA} is a hypothetical DER installation scenario, which represents the physical limit of a distribution network of integrating DERs.

\RV{Although the exact formulation of \eqref{opt:CC-PHCA} does not appear in the literature, it is closely related with almost all proposed methods such as stochastic programming \cite{al-saadi_probabilistic_2017,abad_probabilistic_2018} and robust optimization \cite{wang_distributed_2016}. As pointed out in \cite{geng_data-driven_2019-2}, chance-constrained optimization can be solved using various stochastic programming or robust optimization algorithms, thus \eqref{opt:CC-PHCA}  can serve as a unified problem formulation for hosting capacity analysis under DER uncertainties.}

PHCA is in general challenging to solve for the following reasons \cite{geng_data-driven_2019-2}: (1) the probability distributions of DERs and loads are often not known exactly; (2) even if the exact knowledge of probability distribution is available, it is computationally intractable to accurately evaluate the probability of constraint violation; and (3) chance-constrained programs are generally non-convex, which leads to computational intractability.
\subsubsection{Sample Average Approximation} 
\label{ssub:sample_approximation}
One common solution to deal with the first two difficulties aforementioned is \emph{sample average approximation} \cite{geng_data-driven_2019-2}, which utilizes $N$ i.i.d scenarios $\{ (\bm{\alpha}^i, \bm{d}^i, \bm{e}^i) \}_{i=1}^{N}$ to approximate the violation probability $\epsilon(\bm{\psi})$:
\begin{equation}
\hat{\epsilon}(\bm{\psi}) := 1-\frac{1}{N} \sum_{i=1}^{N} \mathbbm{1} \left( \bm{g}(\bm{\alpha}^i, \bm{d}^i, \bm{e}^i; \bm{\psi} ) \le 0 \right).
\end{equation}
For example, if operational constraints are satisfied in 900 out of 1000 scenarios, then $\hat{\epsilon} = 1-900/1000=0.1$.
With the empirical violation probability $\hat{\epsilon}(\bm{\psi})$, PHCA can be approximated as:
\begin{subequations}
\label{opt:SPHCA}
\begin{align}
\max_{\bm{0} \le \bm{\psi} \le \overline{\bm{\psi}}}~& \bm{1}^\intercal \bm{\psi} \\
\text{s.t.}~& \hat{\epsilon}(\bm{\psi}) \le \overline{\epsilon}
\end{align}
\end{subequations}

Comparing with the original formulation \eqref{opt:CC-PHCA}, \eqref{opt:SPHCA} is relatively easier to solve. It can be reformulated as well-studied optimization problems and solved by commercial optimization solvers, e.g., nonlinear optimization (\ref{ssub:using_nonlinear_optimization}), integer program (Section \ref{ssub:using_mixed_integer_programming}), and Bayesian Optimization (Section \ref{sec:hosting_capacity_maximization_via_bayesian_optimization}).


\subsubsection{Using Nonlinear Optimization} 
\label{ssub:using_nonlinear_optimization}
Formulation \eqref{opt:SPHCA} is a nonlinear program (NLP). The main complexity of solving \eqref{opt:SPHCA} is the complicated nonlinear constraints $\hat{\epsilon}(\bm{\psi}) \le \overline{\epsilon}$. Classical NLP algorithms such as interior point, active set, and sequential quadratic program (sqp) can be applied to solve \eqref{opt:SPHCA}.

\subsubsection{Using Mixed Integer Programming} 
\label{ssub:using_mixed_integer_programming}
Problem \eqref{opt:SPHCA} can be equivalently formulated as a mixed integer program (MIP):
\begin{subequations}
\label{opt:SPHCA-MIP}
\begin{align}
\max_{\bm{\psi}, \bm{z}}~& \bm{1}^\intercal \bm{\psi} \\
\text{s.t.}~& \bm{g}(\bm{\psi}, \bm{\alpha}^{(i)}) \le z_i M \label{opt:relaxed-constr} \\
& \bm{0} \le \bm{\psi} \le \overline{\bm{\psi}} \\
& \frac{1}{N} \sum_{i=1}^{N} z_i \le \overline{\epsilon} \label{opt:violated-constr} \\
& z_i \in \{0,1\},~i=1,2,\cdots,N.
\end{align}
\end{subequations}
where binary variables $z_i = 1- \mathbbm{1} \left( \bm{g}(\bm{\alpha}^i, \bm{d}^i, \bm{e}^i; \bm{\psi} ) \le 0 \right )$. If $z_i = 0$, then all constraints in the $i$th scenario are satisfied \eqref{opt:relaxed-constr}. If $z_i = 1$, then \eqref{opt:relaxed-constr} becomes $\bm{g}(\bm{\alpha}^i, \bm{d}^i, \bm{e}^i; \bm{\psi} ) \le M \bm{1}$. Since $M$ is a large enough coefficient, it is as if all constraints in the $i$th scenario are relaxed. Constraint \eqref{opt:violated-constr} limits the total number of violated scenarios to be no more than accepted bound $\overline{\epsilon}$. Because of the DistFlow equations \eqref{eqn:DistFlow} and line limits \eqref{eqn:hosting_capacity_passive_constraints_line}, \eqref{opt:SPHCA-MIP} is a mixed integer nonlinear program (MINLP).


\section{Probabilistic Hosting Capacity Analysis via Bayesian Optimization} 
\label{sec:hosting_capacity_maximization_via_bayesian_optimization}

\subsection{Introduction to Bayesian Optimization} 
\label{sub:introduction_to_bayesian_optimization}
Bayesian optimization (BayesOpt) is a class of machine-learning-based optimization methods to solve:
\begin{subequations}
\label{opt:BayesOpt}
\begin{align}
\max_{\bm{x}}~& f(\bm{x}) \\
\text{s.t.}~& \bm{x} \in \mathcal{X}
\end{align}
\end{subequations}
in which the feasible set $\mathcal{X}$ is simple (e.g., hyper-rectangle) while the objective $f(\cdot)$ is an expensive black-box derivative-free function \cite{shahriari_taking_2015,frazier_tutorial_2018-1}. More specifically, $f(\bm{x})$ is \emph{expensive} in the sense of evaluating it typically takes a substantial amount of time or bears a monetary cost. Moreover, $f(\cdot)$ could be an unknown function lacking known structural properties like convexity or concavity (black-box). In most cases, information on derivatives is not available either (derivative-free).
BayesOpt aims at find the \emph{global} optimum solution to \eqref{opt:BayesOpt} while using as less function evaluations of $f(\cdot)$ as possible. 

BayesOpt has found successful applications in various areas such as material discovery and medicine \cite{frazier_tutorial_2018-1}. BayesOpt gains its popularity especially due to its success in hyperparameter tuning for machine learning algorithms \cite{shahriari_taking_2015}. Despite the success of BayesOpt in various fields, it has been seldom used in power system studies. Only a handful of papers have incorporated BayesOpt, and all of them used BayesOpt to train machine learning models such as deep neural networks \cite{he_hybrid_2019} or Bayesian Networks \cite{dong_data-driven_2018,zhong_bayesian_2020}.




\subsection{BayesOpt Algorithm} 
\label{sub:bayesopt_algorithm}
BayesOpt has two key components: a Bayesian statistical to model our understanding on the objective function, and an acquisition function for deciding where to sample next. The key steps of BayesOpt algorithms are summarized in Algorithm \ref{alg:bayesopt}.

\begin{algorithm}[H]
\begin{algorithmic}[1]
\STATE input query budget $B$ and initial dataset $\mathcal{D}_0$
\FOR{$i \in \{0,1,2,\cdots,B\}$}
\STATE select the next query $\bm{x}^{i+1}$ by optimizing the acquisition function
\begin{equation}
\bm{x}^{i+1} = \arg \max_{\bm{x} \in \mathcal{X}}~a(\bm{x}; \mathcal{D}_i)
\end{equation}
\STATE query $\bm{x}^{i+1}$ and obtain $f(\bm{x}^{i+1})$ 
\STATE augment dataset $\mathcal{D}_{i+1} = \{\mathcal{D}_i,(\bm{x}^{i+1}, f(\bm{x}^{i+1}))\}$
\STATE update statistical model
\ENDFOR
\end{algorithmic}
\caption{Bayesian Optimization}
\label{alg:bayesopt}
\end{algorithm}
The statistical model, which is often a Gaussian process (GP), provides a Bayesian posterior probability distribution describing potential values for the objective $f(\bm{x})$ of candidate solution $\bm{x}$. The Gaussian process  $\text{GP}(\mu_0, \kappa)$ is a nonparametric model characterized by a mean function $\mu_0: \mathcal{X} \rightarrow \mathbb{R}$ and a kernel function $\kappa: \mathcal{X} \times \mathcal{X} \rightarrow \mathbb{R}$.
The posterior mean and variance evaluated at any point
$\bm{x}$ represent the model’s prediction and uncertainty in the objective function at the point $\bm{x}$.
The mean function $\mu_0$ provides a possible offset. In practice, $\mu_0$ is set to a constant and inferred from data. Common choices of kernel functions include automatic relevance determination (ARD) squared exponential kernel, ARD Matern 3/2 kernel, and ARD Matern 5/2 kernel \cite{snoek_practical_2012}.

The acquisition function determines the mechanism or
policy for selecting the sequence of query points.
Common choices of acquisition functions include expected improvement (EI), probability of improvement (PI), and lower confidence bound \cite{shahriari_taking_2015}. For example, at the $i$th step of Algorithm \ref{alg:bayesopt}, given dataset $\mathcal{D}_i$ and the incumbent solution $\hat{\bm{x}}^i$, the acquisition function $a_{\text{PI}}(\cdot)$ is the probability that a candidate $\bm{x}$ is better than the incumbent solution $\hat{\bm{x}}^i$:
\begin{equation}
a_{\text{PI}}(\bm{x}; \mathcal{D}_i) := \mathbb{P}\left( f(\bm{x}) > f(\hat{\bm{x}}^i) \right) 
\end{equation}
Using similar notations, the acquisition function $a_{\text{EI}}(\cdot)$ is the expected improvement over the incumbent solution $\hat{\bm{x}}^i$:
\begin{equation}
a_{\text{EI}}(\bm{x}; \mathcal{D}_i) := \mathbb{E}\left[ \left( f(\bm{x}) - f(\hat{\bm{x}}^i) \right) \mathbbm{1}(f(\bm{x}) > f(\hat{\bm{x}}^i)) \right]
\end{equation}


\subsection{Probabilistic Hosting Capacity Analysis via BayesOpt} 
\label{sub:hosting_capacity_maximization_via_bayesian_optimization}
BayesOpt solves optimization problems with complicated objective functions, but PHCA \eqref{opt:SPHCA} features a complicated constraint. 
We first reformulate PHCA:
\begin{subequations}
\label{opt:SPHCA-bayesopt}
\begin{align}
\max_{\bm{\psi}}~& c(\bm{\psi}) \\
\text{s.t.}~& \bm{0} \le \bm{\psi} \le \overline{\bm{\psi}}
\end{align}
\end{subequations}
The objective function of \eqref{opt:SPHCA-bayesopt} consists of the original objective $\bm{1}^\intercal \bm{\psi}$ and a penalty term $\rho\left(\max\{\hat{\epsilon}(\bm{\psi} )-\overline{\epsilon},0\} \right)$.
\begin{equation}
c(\bm{\psi}) := \bm{1}^\intercal \bm{\psi} - \rho\left(\max\{\hat{\epsilon}(\bm{\psi} )-\overline{\epsilon},0\} \right).
\end{equation}
When a solution $\bm{\psi}$ is feasible (i.e., $\hat{\epsilon}(\bm{\psi} ) \le \overline{\epsilon}$), the penalty function $\rho\left(\max\{\hat{\epsilon}(\bm{\psi} )-\overline{\epsilon},0\} \right) = 0$ thus the objective function is identical to the original problem. When a solution $\bm{\psi}$ is infeasible ($\hat{\epsilon}(\bm{\psi} ) > \overline{\epsilon}$), then a large penalty $\rho\left(\max\{\hat{\epsilon}(\bm{\psi} )-\overline{\epsilon},0\} \right) \gg \bm{1}^\intercal \bm{\psi}$ is added to the objective, so that any infeasible solutions to \eqref{opt:SPHCA} cannot be optimal to \eqref{opt:SPHCA-bayesopt}. With a carefully chosen penalty function $\rho\left(\max\{\hat{\epsilon}(\bm{\psi} )-\overline{\epsilon},0\} \right)$, the optimal solution to \eqref{opt:SPHCA-bayesopt} is identical to the solution to \eqref{opt:SPHCA}.

\section{Case Study} 
\label{sec:case_study}
We compare BayesOpt with other algorithms on the 56-node South California Edison distribution system \cite{gan_exact_2015}. Full configurations of the distribution system and 365 days of DER and load profiles are available on github\footnote{https://github.com/xb00dx/Distribution-Networks-for-Hosting-Capacity-Analysis}. All numerical simulations were conducted in Matlab R2019a on a laptop with 16GB of RAM and Inter i7-8550U 4-core CPU.


\subsection{Impacts of Problem Formulation} 
\label{sub:formulations}
\RV{
Section \ref{sub:hosting_capacity_maximization} introduces two approaches to solve PHCA problems: using nonlinear optimization and mixed integer programming. We first examine the performances of NLP and MILP algorithms on different formulations. The best approach is chosen to be compared with BayesOpt in Section \ref{sub:main_results}. }

\RV{We use 365 days of scenarios to calculate the empirical violation probability $\hat{\epsilon}(\bm{\psi})$, each day consists of 144 10-min intervals. The complete MINLP formulation \eqref{opt:SPHCA-MIP} has 365 binary variables, 17.7 million\footnote{365 days, 144 snapshot per day, 56 nodes, 6 variables $(\bm{p},\bm{q},\bm{P},\bm{Q},\bm{v}, \bm{l})$ per per node,  $365\times 144\times 56\times 6 \approx 17.7 \times 10^6$.} continuous variables, and 53 million constraints. Formulating such a large-scale optimization problem requires much more RAM than 16GB. Furthermore, it is challenging to directly solve such a large-scale problem using commercial solvers such as CPLEX or GUROBI. Decomposition algorithms such as generalized benders decomposition \cite{geoffrion_generalized_1972} can be applied, but there is no convergence guarantee in general. The NLP formulation \eqref{opt:SPHCA} (and \eqref{opt:SPHCA-bayesopt}) consists of only $|\mathcal{L}|=5$ decision variables and $6$ constraints. The main difficulty of solving \eqref{opt:SPHCA} is constraint $\hat{\epsilon}(\bm{\psi}) \le \overline{\epsilon}$, which involves solving $365 \times 144$ power flow problems thus expensive to evaluate.}

\RV{Comparing with the MINLP formulation \eqref{opt:SPHCA-MIP}, the NLP formulation \eqref{opt:SPHCA} (and \eqref{opt:SPHCA-bayesopt}) is much less demanding on RAM. According to the numerical results in Section \ref{sub:main_results} and Table \ref{tab:fmincon-comp}, NLP is able to solve the problem within in $2\sim 3$ hours, while MINLP algorithms were reported to need $3 \sim 24$ hours to solve similar or smaller problems using carefully designed algorithms \cite{santos_new_2016,xu_enhancing_2019}.}

\RV{Table \ref{tab:fmincon-comp} compares two formulations of PHCA that can be solved by NLP algorithms: (i) the original problem formulation \eqref{opt:SPHCA}; and (ii) reformulating \eqref{opt:SPHCA} using a barrier function. As shown in Table \ref{tab:fmincon-comp}, NLP algorithms on \eqref{opt:SPHCA-bayesopt} were able to find better optimal solutions within the same iteration limits. The average optimality gap is from $19\%$ (interior point) to $49\%$ (active set).}

\begin{table*}[tb]
	\caption{Comparison of Best Incumbent Objectives (better objectives found using \eqref{opt:SPHCA-bayesopt}, compared with using \eqref{opt:SPHCA}) }
	\label{tab:fmincon-comp}
	\centering

	\begin{tabular}{r|ccc|ccc|ccc}
	\hline

	\hline
	  & \multicolumn{3}{c}{\textbf{interior point}} & \multicolumn{3}{|c}{\textbf{sqp}} & \multicolumn{3}{|c}{\textbf{active set} }\\
	\hline
	bestobj  & min & avg  & max & min & avg & max & min & avg & max \\
	\cline{2-10}
	improvement \%  & 0 & 19.0 & 43.1  & 0 & 42.9 & 100 & 0 & 49.1 & 100 \\
	\hline

	\hline
	\end{tabular}
\end{table*}

\subsection{BayesOpt vs NLP Algorithms} 
\label{sub:main_results}
For all algorithms compared in this section, the most time-consuming step is evaluating the empirical violation probability $\hat{\epsilon}(\bm{\psi})$, which involves solving $365 \times 144$ AC power flow problems. With parallel computation on 4 CPU cores, one evaluation of $\hat{\epsilon}(\bm{\psi})$ takes about 5 minutes. Throughout this section, we use the number of $\hat{\epsilon}(\bm{\psi})$ function evaluation calls (nfuncall) as a metric of computation time, e.g., in Fig. \ref{fig:bayesopt_over_others} and Table \ref{tab:bayesopt}.

\subsubsection{Configurations of BayesOpt} 
\label{ssub:configure_bayesopt}
After some trial and error, we found the following objective function for \eqref{opt:SPHCA-bayesopt} works the best numerically:
\begin{equation}
\label{eqn:penalty-objective}
c(\bm{\psi}) = \bm{1}^\intercal \bm{\psi} - |\mathcal{L}| \left(\frac{100 \max\{\hat{\epsilon}(\bm{\psi})-\overline{\epsilon},0\}}{\overline{\epsilon}} \right)^2
\end{equation}
The first coefficient of the penalty term is $\| \partial (\bm{1}^\intercal \bm{\psi}) / \partial \bm{\psi} \|_1 = |\mathcal{L}|$. When a solution $\bm{\psi}$ is infeasible, the squared term in \eqref{eqn:penalty-objective} becomes a large penalty, so that the marginal cost of infeasibility is much larger than the marginal benefit $|\mathcal{L}|$.
We choose the ARD Matern 5/2 kernel for BayesOpt since they are less restrictive and make less assumptions than other kernel functions \cite{snoek_practical_2012}.

\subsubsection{Numerical Results} 
\label{ssub:numerical_results}
\begin{figure}[htbp]
	\centering
	\includegraphics[width=\linewidth]{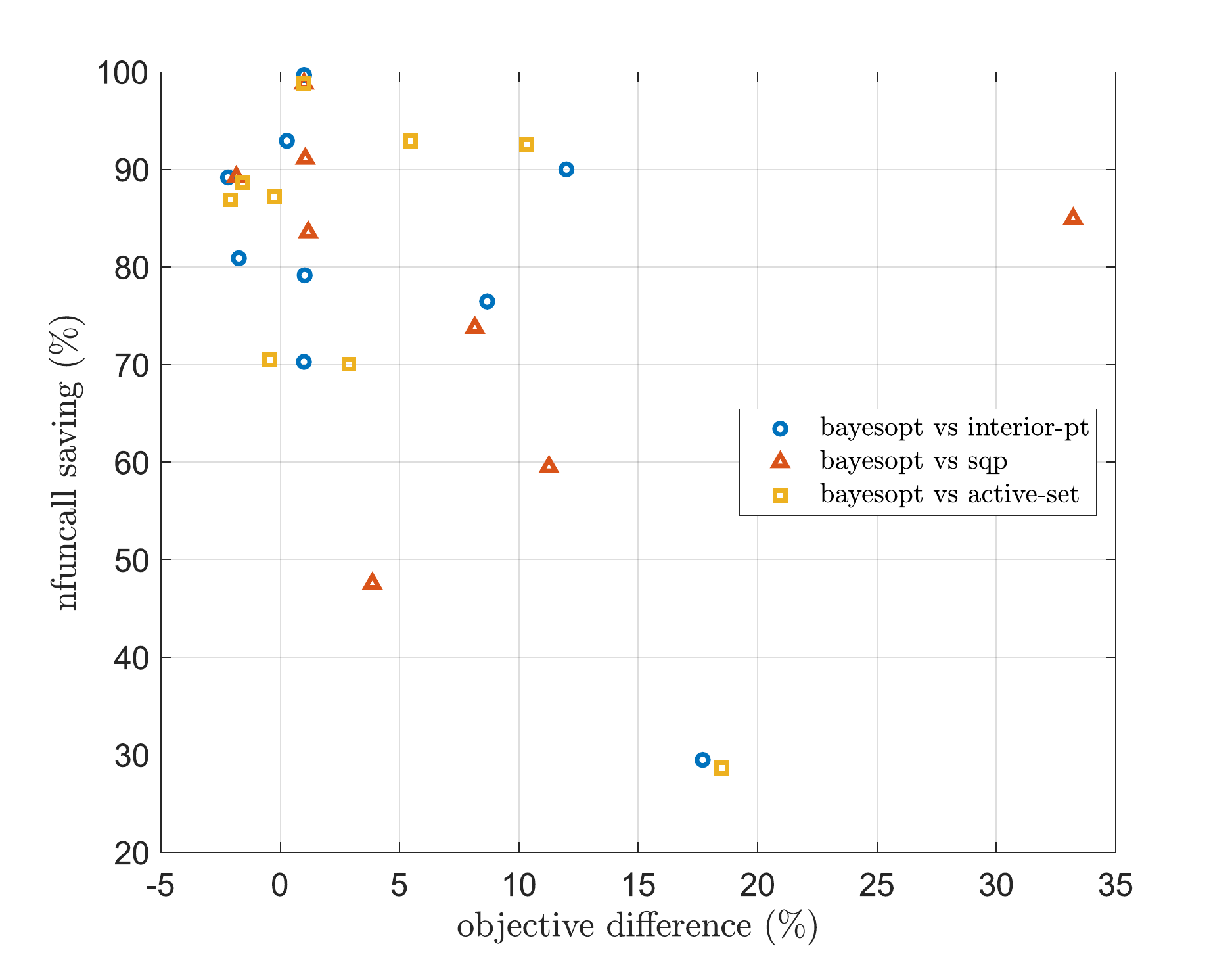}
	\caption{BayesOpt versus other algorithms. Larger x-axis and y-axis values indicate better performance of BayesOpt comparing with the other three algorithms.}
	\label{fig:bayesopt_over_others}
\end{figure}

We solve 10 different PHCA problems (10 experiments), in which we randomly choose $|\mathcal{L}|=5$ different candidate locations  for DERs. In each experiment, we solve the revised PHCA problem \eqref{opt:SPHCA-bayesopt} with objective function \eqref{eqn:penalty-objective} using four algorithms (bayesopt, interior-point, sqp, active-set). Detailed results of 10 experiments are summarized in Table \ref{tab:raw-results}. Main results are in Fig. \ref{fig:bayesopt_over_others} and Table \ref{tab:bayesopt}.

\begin{table*}[tb]
	\caption{Detailed Numerical Results in 10 Experiments}
	\label{tab:raw-results}
	\centering

	\begin{tabular}{l|cccccccccc}
	\hline

	\hline
	 & \multicolumn{10}{c}{ \textbf{bestobj in 10 experiments} } \\
	\hline
	\textbf{no.} & \textbf{1} & \textbf{2} & \textbf{3} & \textbf{4} & \textbf{5} & \textbf{6} & \textbf{7} & \textbf{8} & \textbf{9} & \textbf{10} \\
	\hline
	\textbf{bayesopt}	 &   8.2693 &     7.4855 &     2.7822 &     7.6369 &     5.8029 &     3.6995 &     9.1246 &     7.8330 &     9.2312 &     3.2766  \\
	\textbf{interior point}	 &     8.4539 &     6.8883 &     2.3638 &     7.5594 &     5.9050 &     3.6628 &     9.0989 &     7.7559 &     3.0390 &     2.9258  \\
	\textbf{sqp}	 &     8.1732 &    7.6258 &         0 &    7.0610 &    5.2156 &    3.6628 &    9.0295 &    7.5418 &    2.8051 &    2.4595 \\
	\textbf{active set}	 &     8.4444 &     7.6057 &     2.3478 &     7.6559 &     5.6404 &     3.6628 &     8.6514 &     7.8681 &     2.9478 &     2.9698  \\
	\hline

	\hline
 	& \multicolumn{10}{c}{ \textbf{nfuncall in 10 experiments} } \\
	\hline	
	\textbf{no.} & \textbf{1} & \textbf{2} & \textbf{3} & \textbf{4} & \textbf{5} & \textbf{6} & \textbf{7} & \textbf{8} & \textbf{9} & \textbf{10} \\
	\hline
	\textbf{bayesopt}	 &       37 &     40 &    122 &     59 &     68 &      2 &     20 &    118 &    126 &     17  \\
	\textbf{interior point}	 &    342 &   170 &   173 &   283 &   356 &   617 &   283 &   397 &   114 &   170 \\
	\textbf{sqp}	 &    225 &    371 &     56 &    225 &    168 &    168 &    225 &    225 &    113 &    113 \\
	\textbf{active set}	 &    282 &    352 &    171 &    461 &    227 &    168 &    283 &    400 &    171 &    228  \\
	\hline

	\hline	
	\end{tabular}
\end{table*}

\begin{figure*}	
\centering
\begin{subfigure}[t]{0.49\linewidth}
	\centering
	\includegraphics[width=\linewidth]{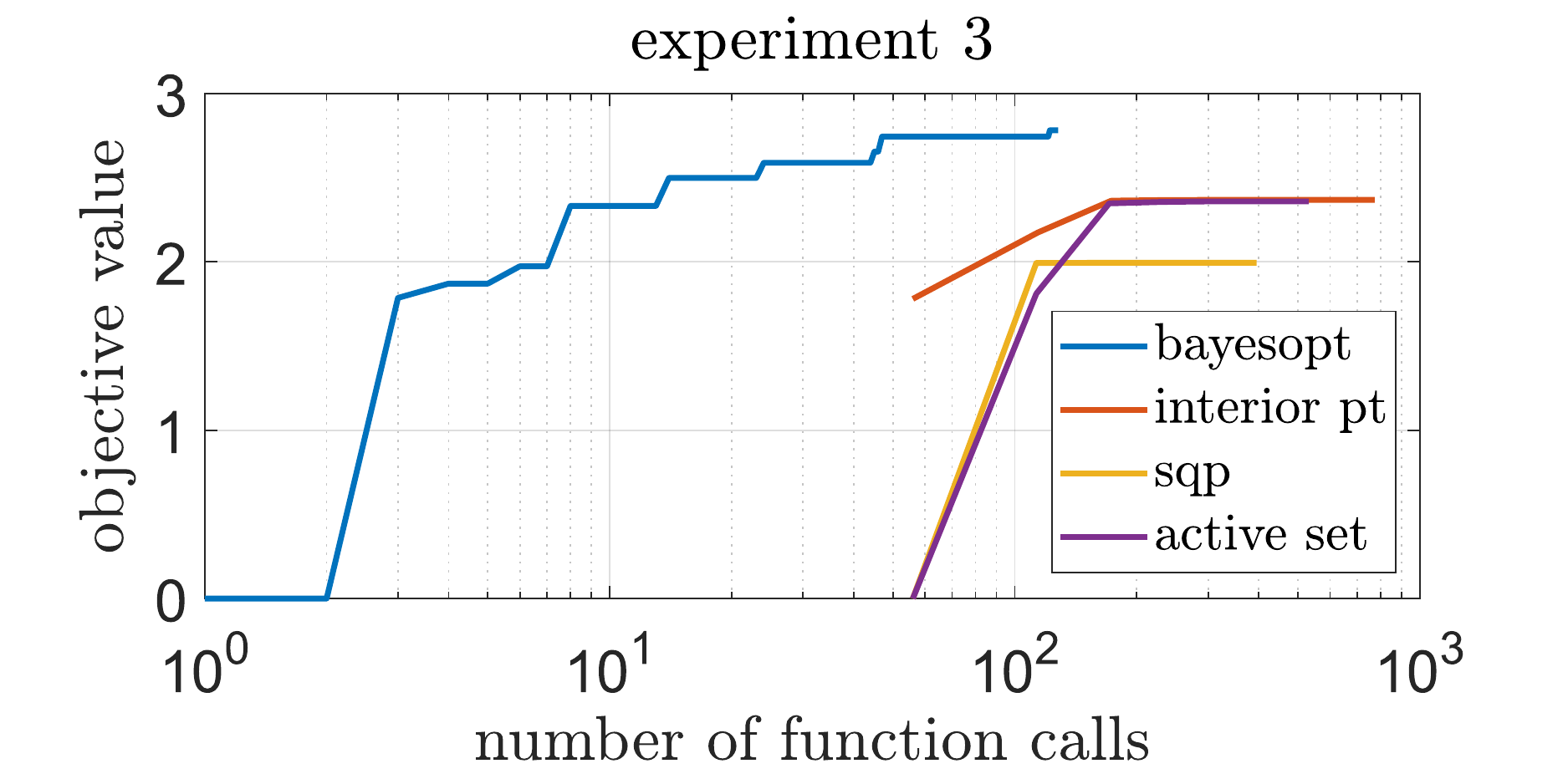}
	\caption{Experiment 3: best incumbent objectives}\label{fig:objhistory-experiment3}		
\end{subfigure}
\begin{subfigure}[t]{0.49\linewidth}
	\centering
	\includegraphics[width=\linewidth]{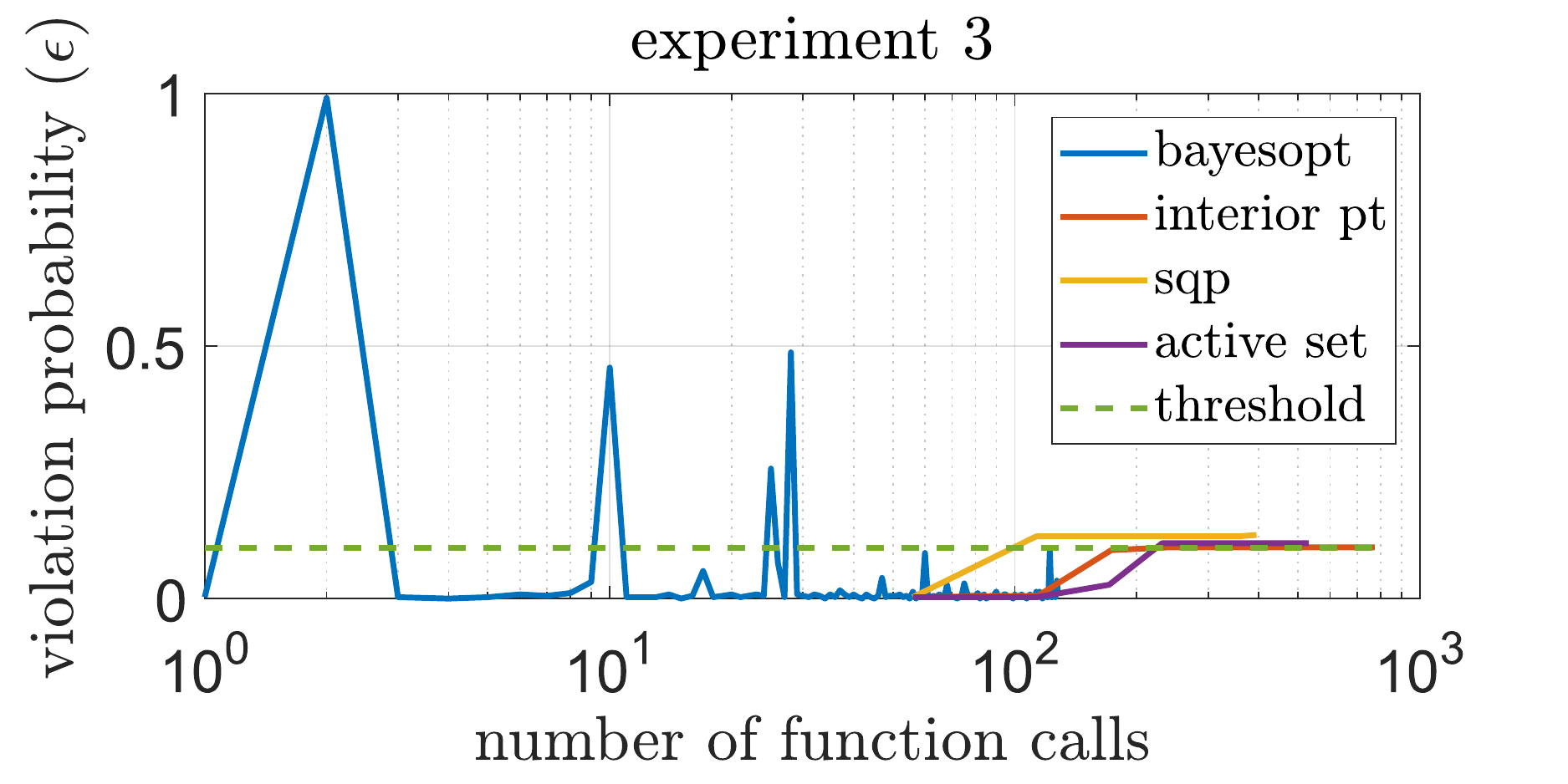}
	\caption{Experiment 3: violation probabilities}\label{fig:probhistory-experiment3}		
\end{subfigure}
\begin{subfigure}[t]{0.49\linewidth}
	\centering
	\includegraphics[width=\linewidth]{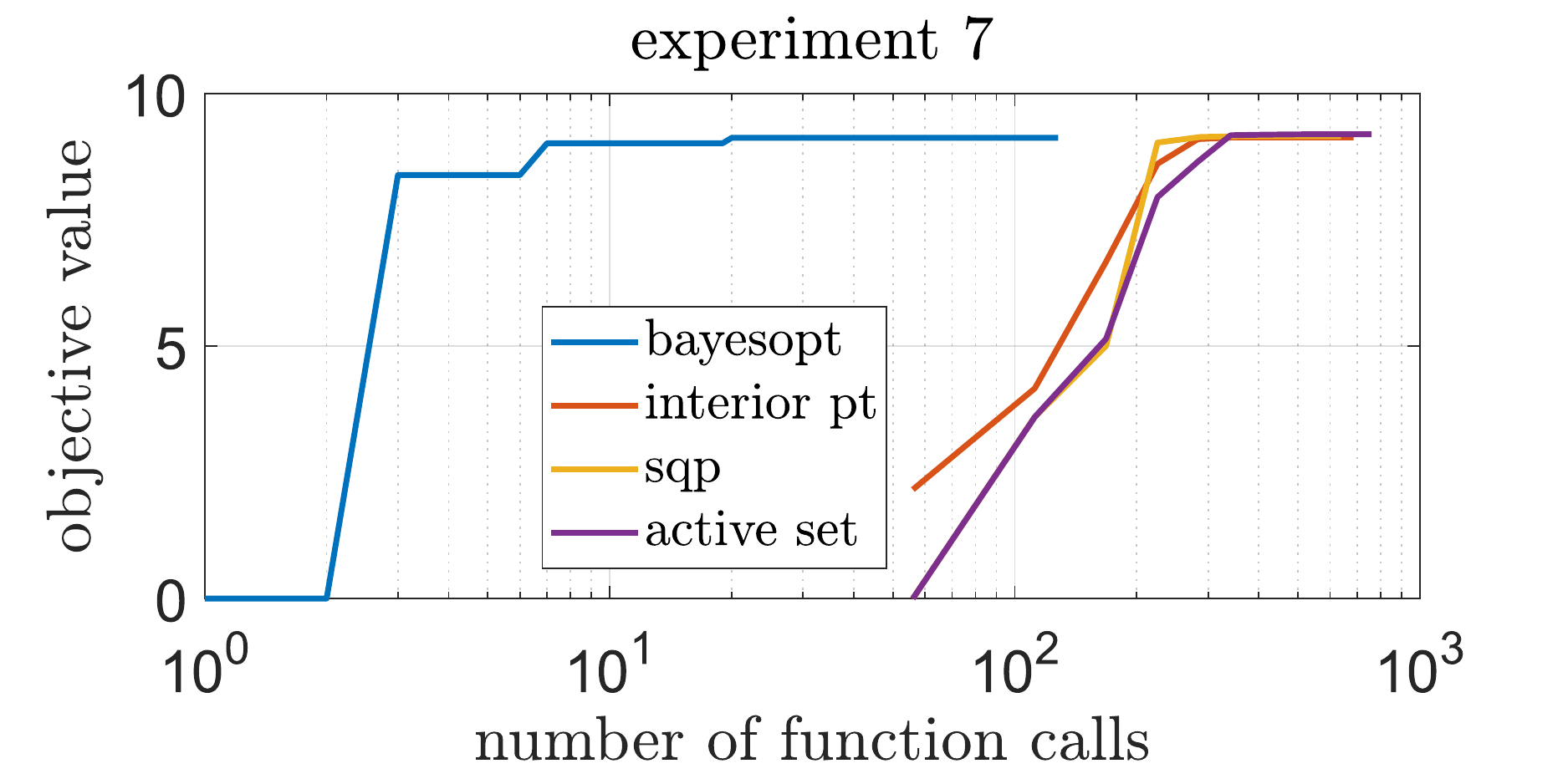}
	\caption{Experiment 7: best incumbent objectives}\label{fig:objhistory-experiment7}
\end{subfigure}
\begin{subfigure}[t]{0.49\linewidth}
	\centering
	\includegraphics[width=\linewidth]{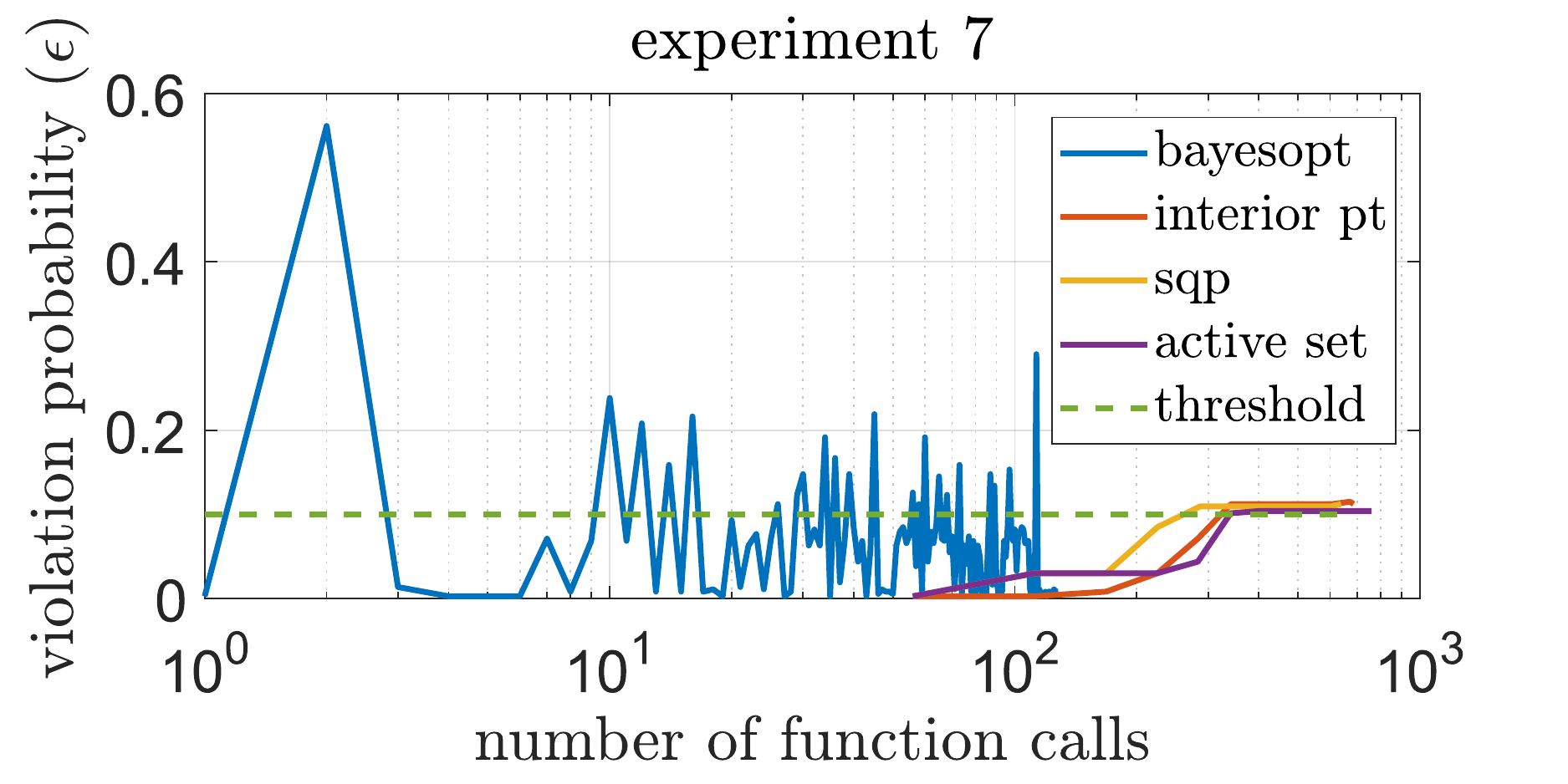}
	\caption{Experiment 7: violation probabilities}\label{fig:probhistory-experiment7}
\end{subfigure}
\begin{subfigure}[t]{0.49\linewidth}
	\centering
	\includegraphics[width=\linewidth]{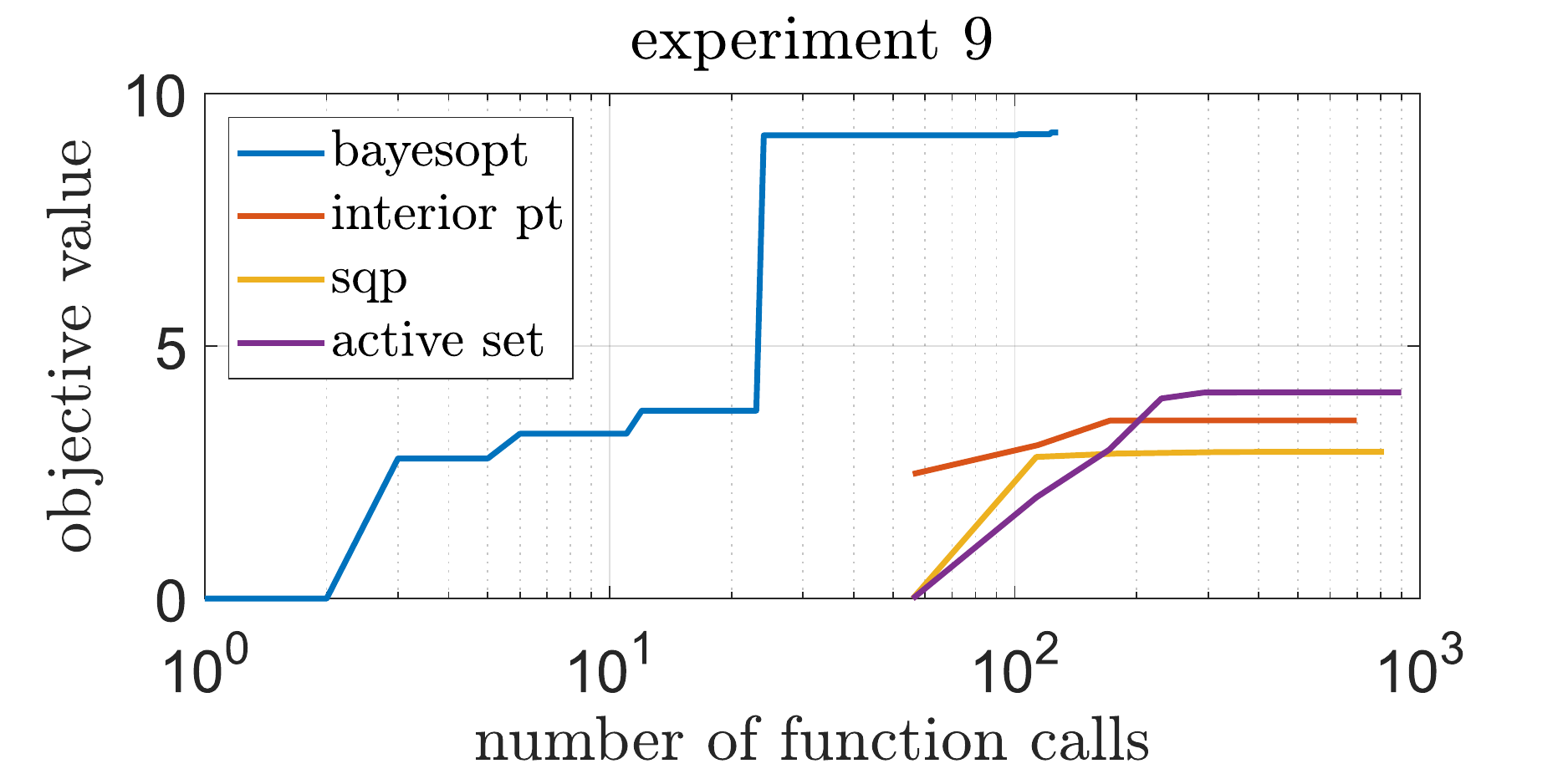}
	\caption{Experiment 9: best incumbent objectives}\label{fig:objhistory-experiment9}
\end{subfigure}
\begin{subfigure}[t]{0.49\linewidth}
	\centering
	\includegraphics[width=\linewidth]{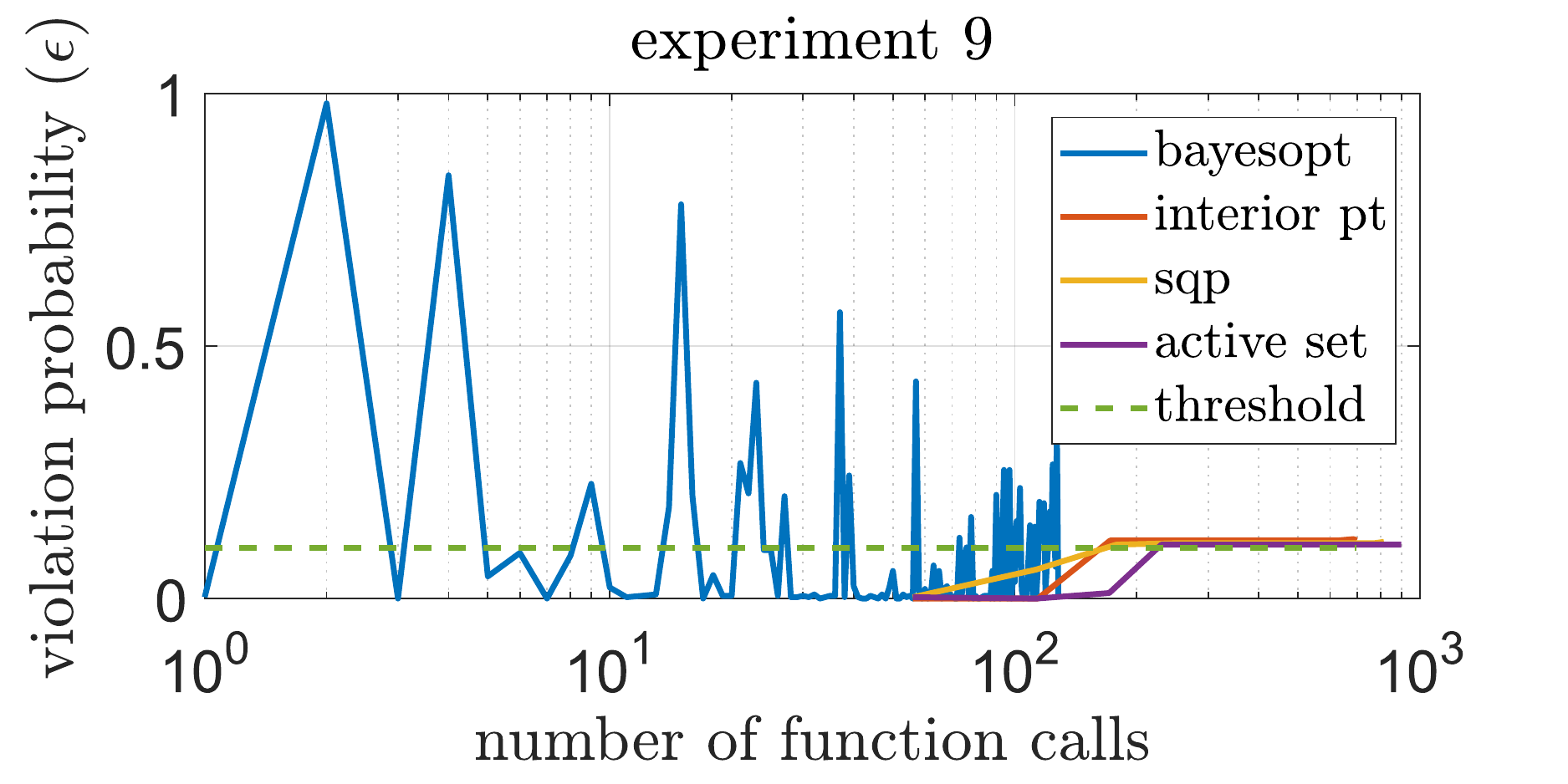}
	\caption{Experiment 9: violation probabilities}\label{fig:probhistory-experiment9}
\end{subfigure}
\caption{Details of violation probabilities and objective values on three experiments: experiment 3 (sqp failed completely), experiment 7 (NLP algorithms found slightly better solutions than BayesOpt), experiment 9 (bayesopt significantly outperforms NLP algorithms).}\label{fig:3experiments}
\end{figure*}

Fig. \ref{fig:bayesopt_over_others} compares the performance of BayesOpt with the other three algorithms (blue circles: interior point; red triangles: sqp; yellow squares: active-set). 
The x-axis of Fig. \ref{fig:bayesopt_over_others} denotes the improvement of BayesOpt over other three algorithms in terms of better objectives. Since PHCA is a maximization problem, larger x-axis values indicate better solutions found by BayesOpt, and negative values indicate other algorithms found better solutions than BayesOpt. The y-axis represents the improvement of BayesOpt over other algorithms in terms of computational time, which is quantified by the number of function calls of $\hat{\epsilon}(\bm{\psi})$ (nfuncall).
Larger y-axis values indicate less computation time of BayesOpt, e.g., $70\%$ means that BayesOpt utilized 70\% \emph{less} nfuncalls than other algorithms. In most cases, BayesOpt found solutions that are better than other algorithms while using much less evaluations of $\hat{\epsilon}(\bm{\psi})$. In $3\sim4$ cases (top-left corner of Fig. \ref{fig:bayesopt_over_others}), NLP algorithms found slightly better solutions than PHCA (optimality gaps $0.1\%\sim 2.5\%$) using much longer computational time. One such example is experiment 7 in Figs. \ref{fig:objhistory-experiment7} and \ref{fig:probhistory-experiment7}.

It is worth mentioning that only 8 experiments (24 points) are plotted in Fig. \ref{fig:bayesopt_over_others}. This is because BayesOpt performs significantly better than the other three algorithms in experiments 3 and 9. In experiment 9 (shown in Figs. \ref{fig:objhistory-experiment9} and \ref{fig:probhistory-experiment9}), BayesOpt found an optimal solution with objective value $9.23$, while other algorithms were stuck at a local solution with objective $2.80\sim 3.03$, this contributes to the $200+\%$ improvements of best incumbent objective (bestobj) in Table \ref{tab:bayesopt}. In experiment 3 (shown in Figs. \ref{fig:objhistory-experiment3} and \ref{fig:probhistory-experiment3}), sqp failed to find a better solution than initial point within iteration limit. This causes the ``inf'' in max objective gaps. Values in brackets of columns ``sqp'' are calculated including experiment 3.



\begin{table*}[tb]
	\caption{Compare BayesOpt with others, higher values indicate BayesOpt perform better}
	\label{tab:bayesopt}
	\centering

	\begin{tabular}{l|ccc|ccc|ccc}
	\hline

	\hline
	\multicolumn{1}{c}{bayesopt vs other} & \multicolumn{3}{|c}{\textbf{interior point}} & \multicolumn{3}{|c}{\textbf{sqp}} & \multicolumn{3}{|c}{\textbf{active set} }\\
	\cline{2-10}
	\multicolumn{1}{c|}{algorithms} &  min & avg  & max & min & avg & max & min & avg & max \\
	\hline
	improvement in bestobj  & -2.2\% & 24.2\% & 203.8\% & -1.8\% & 31.9\% & 229.1\% (inf)  & -2.1\% & 24.7\% & 213.2\% \\	
	improvement in nfuncall & -10.5\%  & 69.8\% & 99.7\% & -11.5\% (-117.9\%) & 68.6\% (49.9\%) & 98.8\%  & 26.3\% & 74.3\% & 98.8\% \\
	\hline

	\hline
	\end{tabular}
\end{table*}

\subsubsection{Why BayesOpt Outperforms Other Algorithms} 
\label{ssub:why_bayesopt_outperforms_other_algorithms}
\RV{Fig. \ref{fig:bayesopt_over_others} and Table \ref{tab:bayesopt} show that the proposed BayesOpt approach is able to find better solutions (25\% higher hosting capacity) with 70\% savings in computation time. This is mainly due to the following two reasons. First, BayesOpt \emph{explicitly} considers the fact that evaluating $c(\bm{\psi})$ is expensive. In numerical simulations, we observe that NLP algorithms usually require several calls to the objective function $c(\bm{\psi})$ in each iteration, thus they are significantly slower than BayesOpt. Second, NLP algorithms explore the feasible region in a continuous manner (continuous objective improvements in Fig. \ref{fig:3experiments}), e.g., the central path of interior-point algorithm, thus they might be stuck at suboptimal optimal solutions. By contrast, BayesOpt explores the entire feasible region, balances the tradeoff between exploration and exploitation thus seeks the global optimal solution.}

\subsection{Extensions} 
\label{sub:extensions}
\RV{The proposed framework can be easily extended towards more complicated models such as three-phase unbalanced power flow and mesh distribution networks. Many existing techniques can be easily integrated into the proposed framework as well. For example, evaluating the expensive constraint can be accelerated via parallel computation and multi-parametric programming \cite{taheri_fast_2020}. }

\section{Concluding Remarks} 
\label{sec:concluding_remarks}
We study the probabilistic hosting capacity analysis problem in distribution networks with significant uncertainties from distributed energy resources (DERs) and residential loads. PHCA is often formulated as large-scale non-convex optimization problems. To address the core computational challenges of PHCA, we propose a computational framework based on Bayesian Optimization (BayesOpt). Comparing with nonlinear optimization algorithms, the proposed BayesOpt approach returns better optimal solutions while using much less computation time on average.




\bibliographystyle{IEEEtran}
\bibliography{references}

\end{document}